\documentclass[conference]{IEEEtran}

\usepackage{amsmath,amsfonts,amssymb}
\usepackage[utf8]{inputenc}
\usepackage{graphicx}

\usepackage{algorithmicx}
\usepackage{algpseudocode}
\usepackage{algorithm}

\newtheorem{definition}{Definition}

\newcommand{\WRP}{\par\qquad\enspace}

\algnewcommand{\IIf}[1]{\State\algorithmicif\ #1\ \algorithmicthen}
\algnewcommand{\EndIIf}{\unskip\ \algorithmicend\ \algorithmicif}

\begin{document}

\title{An Improved SCFlip Decoder for Polar Codes}

\author{\IEEEauthorblockN{L.Chandesris\IEEEauthorrefmark{2}\IEEEauthorrefmark{3},
V.Savin\IEEEauthorrefmark{2}, D.Declercq\IEEEauthorrefmark{3}  \\ {ludovic.chandesris@cea.fr,  valentin.savin@cea.fr,  declercq@ensea.fr}}
\IEEEauthorrefmark{2}CEA-LETI / Minatec, Grenoble, France \quad
\IEEEauthorrefmark{3}ETIS, ENSEA/UCP/CNRS, Cergy-Pontoise, France}

\maketitle

\begin{abstract}
This paper focuses on the recently introduced Successive Cancellation Flip (SCFlip) decoder of polar codes. Our contribution is twofold. First, we propose the use of an optimized metric to determine the flipping positions within the SCFlip decoder, which improves its ability to find the first error that occurred during the initial SC decoding attempt. We also show that the proposed metric allows closely approaching the performance of an ideal SCFlip decoder. Second, we introduce a generalisation of the SCFlip decoder to a number of $\omega$ nested flips, denoted by SCFlip-$\omega$, using a similar optimized metric to determine the positions of the nested flips. We show that the SCFlip-2 decoder yields significant gains in terms of decoding performance and competes with the performance of the CRC-aided SC-List decoder with list size $L=4$, while having an average decoding complexity similar to that of the standard SC decoding, at medium to high signal to noise ratio.   

\end{abstract}



\section{Introduction}\label{sec:num1}

Polar codes are a new class of error-correcting codes, proposed by Arikan in \cite{arikan2009channel}, which provably achieve the capacity of any symmetric binary-input memoryless channel under successive cancellation (SC) decoding. However, for short to moderate blocklengths, the frame error rate (FER) performance of polar codes under successive cancellation decoding does not compete with other families of codes such as LDPC or Turbo-Codes.

In \cite{tal2011list}, a Successive Cancellation list-decoding (SCL) has been proposed, which significantly outperforms the simple SC decoding, and approaches the Maximum-Likelihood (ML) performance at high signal to noise ratio (SNR). Moreover, when applied to polar codes concatenated with an outer cyclic redundancy check (CRC) code -used to identify the correct message from the decoded list- it has been shown that the SCL decoder may successfully compete with other families of capacity approaching codes, like Low Density Parity Check (LDPC) codes. However, SCL decoder suffers from high storage and computational complexity, which grows linearly with the size of the list. Several improvements have been proposed to reduce its computational complexity, such as Stack decoding (SCS) in \cite{niu2012stack}, but at a cost of an increasing storage complexity.

Successive Cancellation Flip decoder has been introduced in \cite{bastani2012polar} for the BEC channel and generalised to binary-input additive white Gaussian noise (BI-AWGN) channel by using a CRC in \cite{afisiadis2014low}. It is close to the order statistic decoding proposed in \cite{fossorier1995soft} and which has been specifically used for polar codes in \cite{wu2016ordered}. The idea is to allow a given number of new decoding attempts, in case that a failure of the initial SC decoding attempt is detected by the CRC. Each new attempt consists in flipping one single decision - starting with the least reliable one, according to the absolute value of the log-likelihood ratio (LLR) - of the initial SC attempt, then decoding the subsequent positions by using standard SC decoding. The above procedure is iterated until the CRC is verified or a predetermined maximum number of flips is reached. The SCFlip decoder provides an interesting trade-off between decoding performance and decoding complexity, since each new decoding attempt is only performed if the previous one failed. Consequently, the average computational complexity of the SCFlip decoder approaches the one of SC decoder at medium to high SNR, while competing with the CRC-aided SCL with $L$ = 2, in terms of error correction performance. 
	
In this work we propose two improvements to the SCFlip decoder, aimed at both increasing the error correction performance and reducing the computational complexity. First, we propose the use of a new metric to determine the flipping positions within the SCFlip decoder. The proposed metric takes into account the sequential aspect of the SC decoder, and we show it yields an improved FER performance and a reduced computational complexity compared to LLR-based metric used in \cite{afisiadis2014low}. Second, we introduce a generalization of the SCFlip decoder to a number of $\omega$ nested flips, denoted by SCFlip-$\omega$. We show that the SCFlip-2 decoder with the proposed metric to select the two flipping positions competes with the CRC-aided SCL decoder with $L = 4$, in terms of decoding performance, while having an average decoding complexity similar to that of the standard SC decoding at medium to high SNR. Furthermore, we also use an \textit{Oracle-assisted} decoder as in \cite{afisiadis2014low} to determine the lower bound of these SCFlip decoders and shows that both proposed algorithms for $\omega=1$ and $\omega=2$ can closely approach the optimal performance. 
	
The paper is organised as follows. In Section \ref{sec:num2}, a review of Polar Codes is presented. Section \ref{sec:num3} describes the SCFlip-$\omega$ decoder. Section \ref{sec:num4} presents the proposed metric for the selection of the flipping positions. Simulation results are presented in section \ref{sec:num5}.

\section{Preliminaries}\label{sec:num2}

\subsection{Polar Codes}

A Polar Code is characterized by the three-tuple $(N,K,\mathcal{I})$, where $N=2^n$ is the blocklength, $K$ the number of information bits and $\mathcal{I}$ is the set of $K$ indices indicating the position of the information bits inside the block of size $N$. Bits corresponding to positions $i\not\in {\cal I}$ are referred to as \textit{frozen bits} and are fixed to pre-determined values known at both the encoder and the decoder. 

We denote $\textbf{U}=u_0^{N-1}$ the data vector, of length $N$, containing $K$ information bits at the positions $i\in I$, and $N-K$ frozen bits that are set to zero. The encoded vector, denoted by $\mathbf{X}$, is obtained by:
$$\mathbf{X}=\textbf{U} \cdot \mathbf{G}_{N}$$
where $\mathbf{G}_N$ is the generator matrix defined as in \cite{arikan2009channel}. We further denote by $\mathbf{Y}$ the data received from the channel and used at the decoder input. $\mathbf{\hat{U}}=\hat{u}_0^{N-1}$ denotes the decoder's output, with $\hat{u}_i$ being the estimation of the bit $u_i$.

\subsection{Successive Cancellation Decoder}

SC decoder is the standard low complexity decoder of polar codes given in \cite{arikan2009channel}. The decoding process consists in taking a decision on bit $u_i$, denoted $\hat{u_i}$, according to the sign of the LLR: 
\begin{equation}
\text{L}(u_i)=\log \left(\frac{\Pr(u_i=0 | \mathbf{Y}, \hat{u}_0^{i-1})}{\Pr(u_i=1 | \mathbf{Y}, \hat{u}_0^{i-1})} \right)$$

\noindent using the decision function $h$:
$$
\hat{u}_i = h(L(u_i)) \stackrel{\text{def}}{=}  \left\{
    \begin{array}{ll}
        \qquad u_i & \mbox{if } i \notin \mathcal{I} \\
        \frac{1 - \text{sign}(L(u_i))}{2} & \mbox{if } i \in \mathcal{I}

    \end{array}
\right.
\end{equation}

\noindent where by convention $\text{sign}(0) = \pm 1$ with equal probability. Note that the computations and decisions are performed sequentially in SC decoder, as the estimation of the current bit $u_i$ depends on the previous decoded bits $\hat{u}_0^{i-1}$. 

\section{Definition and Analysis of Scflip-$\omega$ Decoders}\label{sec:num3}

\subsection{Definition of SCFlip-$\omega$ Decoders}\label{sec:num6}

Let $\mathcal{C}(N,K+r,\mathcal{I})$ denotes the serial concatenation of an outer $(K+r, K)$ CRC code and an inner $(N, K+r, \mathcal{I})$ Polar code. Note that the number of unfrozen positions of the Polar code is $K+r$, where $K$ is the number of information bits, and $r$ is the size of the  CRC. 

The SCFlip decoder \cite{afisiadis2014low} consists of a standard SC decoding, possibly followed by maximum number $T$ of new decoding attempts, until no errors are detected by the CRC check. Each new decoding consists of (i) flipping only one decision of the initial SC attempt, then (ii) decoding the subsequent positions by using standard SC decoding. The $T$ {\em flipping positions} are those corresponding to the lowest absolute values of the LLRs computed during the initial SC attempt. 

\begin{algorithm}[!tbh]
\caption{SCFlip decoder with $\omega=1$}
\label{SCF1}
\begin{algorithmic}[1]
\Procedure{SCFlip-1}{$\mathbf{Y},\mathcal{I},T^{(1)}$}
	\State ($\hat{u}_0^{N-1},\{L(u_i)\}_{i \in \mathcal{I}}) \leftarrow $SC$(\mathbf{Y},\mathcal{I},\varnothing)$ 	
	\IIf{CRC($\hat{u}_0^{N-1}$)=success} Return $\hat{u}_0^{N-1}$ \EndIIf
	\State $\mathcal{L}_{\text{flip}} =$ FlipDetermine($\{L(u_i)\}_{i \in \mathcal{I}}$,$-1$,$T^{(1)}$)
	\For {$j=0,..,T^{(1)}-1$}	
	 	\State $\hat{u}_0^{N-1} \leftarrow $SC$(\mathbf{Y},\mathcal{I},\mathcal{L}_{\text{flip}}(j)$) 
		\IIf{CRC($\hat{u}_0^{N-1}$)=success} Return $\hat{u}_0^{N-1}$ \EndIIf 
	 \EndFor
	\State Return $\hat{u}_0^{N-1}$
\EndProcedure
\end{algorithmic}
\end{algorithm}

However, the success of the SCFlip decoding depends on (i) the ability to find the very first error that occurred during the initial SC attempt, and (ii) the ability of SC to successfully decode the subsequent positions, once the first position in error has been flipped. In this work we introduce two new enhancements to the SCFlip decoder, aimed at improving the two above-mentioned characteristics. We propose the use of an optimized metric to determine the flipping positions, building upon the probability of a given position being first error that occurred in the initial SC attempt (see section IV). We note that the global structure of the SCFlip decoding stays the same, the only difference being on the generation of the ordered list of flipping positions, denoted by $\mathcal{L}_{\text{flip}}$. We also introduce a generalisation of the SCFlip decoder to a number of $\omega$ nested flips: the first flip is performed on one decision of the initial SC attempt, while the $i$-th flip ($2\leq i \leq \omega$) is performed on a bit position belonging to the new decoding trajectory determined by the previous flips ($1$ to $i-1$). Such a sequence of $\omega$ nested flips will also be referred to as an {\em order-$\omega$ flip}. The SCFlip-$\omega$ decoding for $\omega=1$ and $\omega=2$ are presented in Algorithm~1 and Algorithm~2, respectively. 

\begin{algorithm}[!tb]
\caption{SCFlip decoder with $\omega=2$}
\label{SCF2}
\begin{algorithmic}[1]
\Procedure{SCFlip-2}{$\mathbf{Y},\mathcal{I},T^{(1)},\{T^{(2,1)},T^{(2,2)}\}$}
	\State ($\hat{u}_0^{N-1},\{L(u_i)\}_{i \in \mathcal{I}}) \leftarrow $SC$(\mathbf{Y},\mathcal{I},\varnothing)$ 	
	\IIf{CRC($\hat{u}_0^{N-1}$)=success} Return $\hat{u}_0^{N-1}$ \EndIIf 
	\State $\mathcal{L}_{\text{flip}}^{(1)} =$ FlipDetermine($\{L(u_i)\}_{i \in \mathcal{I}}$,$-1$,$T^{(1)}$)
	\For {$j=0,..,T^{(1)}-1$}	
	 	\State ($\hat{u}_0^{N-1},\{L(u_i)\}_{\mathcal{I}}) \leftarrow $SC$(\mathbf{Y},\mathcal{I},\mathcal{L}_{\text{flip}}^{(1)}(j)$) 
		\IIf{CRC($\hat{u}_0^{N-1}$)=success} Return $\hat{u}_0^{N-1}$ \EndIIf 
			\If {$j < T^{(2,1)}$}
				\State $\mathcal{L}_{\text{flip}}^{(2)} =(\mathcal{L}_{\text{flip}}^{(2)},$ FlipDetermine ( \WRP \hspace{2cm}$\{L(u_i)\}_{\mathcal{I}}, j,\mathcal{L}_{\text{flip}}^{(1)}(j),T^{(2,2)}$ )
			\EndIf				 
	 \EndFor

		\For {$i=0,..,T^{(2,1)}-1$}	
		 	\For {$j=0,..,T^{(2,2)}-1$}	
		 	\State $\hat {u}_0^{N-1} \leftarrow $SC$ ( \mathbf{Y},\mathcal{I}, $\WRP \hspace{2cm} $\{\mathcal{L}_{\text{flip}}^{(1)}(i),\mathcal{L}_{\text{flip}}^{(2)}(j+i \cdot T^{(2,2)})\}$) 
			\If {CRC($\hat{u}_0^{N-1}$)=success}
				\State Return $\hat{u}_0^{N-1}$
			\EndIf	
		 \EndFor
		 \EndFor
	\State Return $\hat{u}_0^{N-1}$
\EndProcedure
\end{algorithmic}
\end{algorithm}

A simple and efficient implementation of SCFlip-$\omega$ decoder consists in incrementing $\omega$ recursively. As long as no errors are detected by the CRC code, we proceed step by step from SC decoder, SCFlip with 1 flip for a number of $T^{(1)}$ attempts, then SCFlip with 2 flips for a number of $T^{(2)}$ attempts. New decoding attempts in SCFlip decoder are similar to the standard SC decoder, the only difference being on the hard decision function. Thus, we use the notation SC($\mathbf{Y},\mathcal{I},\mathcal{E}$), where $\mathcal{E}$ is a set of $\omega$ indices corresponding to the flipping positions. The hard decision function $h'$ can be defined as follow:
$$
\hat{u}_i = h'(L(u_i)) \stackrel{\text{def}}{=} \left\{
    \begin{array}{ll}
        h(L(u_i)) & \mbox{if } i \notin \mathcal{E} \\
        1-h(L(u_i)) & \mbox{if } i \in \mathcal{E} 
    \end{array}
\right.
$$

Note that we have a standard SC decoder if $\mathcal{E}=\varnothing$. In Algorithm~\ref{SCF1} and Algorithm~\ref{SCF2}, candidate positions for the $\omega$-flips are stored in an ordered list denoted by $\mathcal{L}_{\text{flip}}^{(\omega)}$ of size $T^{(\omega)}$ and generated by the function \textit{FlipDetermine}, described in Algorithm \ref{FD2}, which is performed using a metric denoted $M$. The calculation of this metric will be discussed in section IV.

Algorithm \ref{FD2} is used in SCFlip-1 as well as SCFlip-2 to determine $T$ indexes strictly larger than $k_1$ of the least reliable decisions according to the proposed metric. To do so, we first calculate the metric vector $m$, then generate an index vector $J$, such that $m(J)$ is sorted in descending order (this operation is performed by the function denoted {\em sort\_index}). For SCFlip-1, $\mathcal{L}_{\text{flip}}^{(1)}$ is generated by calling this function once with $T=T^{(1)}$ and $k_1=-1$. For the SCFlip-2, we have two degrees of freedom concerning the choice of the first and second flipping position. Therefore, we use two parameters $T^{(2,1)}$ and $T^{(2,2)}$ to characterize the flips of order 2. $T^{(2,1)}$ is the number of positions from the $\mathcal{L}_{\text{flip}}^{(1)}$ list, for which flips of order-2 will be explored. The number of order-2 flips explored for each of these positions is given by $T^{(2,2)}$. The ordered list $\mathcal{L}_{\text{flip}}^{(2)}$ of size $T^{(2)}$ is obtained by concatenating the vectors of size $T^{(2,2)}$ returned by $T^{(2,1)}$ successive calls to this function. For each call, $k_1$ corresponds to the position of the first bit flipped. The maximum number of order-2 attempts is $T^{(2)}=T^{(2,1)} \cdot T^{(2,2)}$.

\begin{algorithm}[!t]
\caption{FlipDetermine}
\label{FD2}
\begin{algorithmic}[1]
\Procedure{FlipDetermine}{$\{L(u_i)\}_{i \in \mathcal{I}},k_1,T$}
	\For {$i = 0,\dots,N-1$}
		\If{$i \in \mathcal{I}$ and $i>k_1$} $m(i)=M_{\alpha}(u_i)$\Else \text{ }$m(i)=0$ \EndIf
	\EndFor
	\State J $\leftarrow$ sort\_index($m$)
	\State Return J$(0:T-1)$
\EndProcedure
\end{algorithmic}
\end{algorithm}

\subsection{Oracle-Assisted Decoder and Order of a Noise Realization}

Following \cite{afisiadis2014low}, we distinguish between \textit{channel-generated errors} (CGE) and \textit{propagation errors} (PE) in the SC decoding. \textit{Propagation errors} are generated by an erroneous decision, which propagates in the decoding process, while \textit{channel-generated errors} correspond to erroneous decisions which are only generated by the noise realization at the decoder’s input. From these definitions, the first error in SC decoding is necessary a CGE. 
	
	In \cite{afisiadis2014low}, an Oracle-assisted decoder (OA-SC) has been introduced to count the number $\omega$ of \textit{channel-generated errors} which occur during a SC decoding. OA-SC performs a standard SC decoder with a hard decision function $h^{(OA)}$ modified to ensure that the decision is correct and no error will propagate during the process: $h^{(OA)}(L(u_i))=u_i$. Hence, $\omega$ is defined by:
$$\omega= \underset{i \in \mathcal{I}}{\#} \{h^{(OA)}(L(u_i)) \neq h(L(u_i) \},$$
	
\noindent where the symbol $\#\{A\}$ denotes the number of times the condition $A$ is verified. In this paper, the parameter $\omega$ is referred to as {\em order of a noise realization}.

Note that we use the same notation $\omega$ for the flip order in the SCFlip-$\omega$ decoder and the \textit{order of a noise realization} as they are directly related. Indeed, the SCFlip-$\omega$ is able to decode a noise realization of order $\omega' \leq \omega$, provided that (i) the corresponding order-$\omega'$ flip has been selected in the corresponding ordered list $\mathcal{L}_{\text{flip}}^{(\omega')}$ and (ii) the CRC is not verified by one of the previous decoding attempts. 
As in \cite{afisiadis2014low}, we use the OA-SC decoder to predict the optimal performance of a SCFlip-$\omega$ decoder, regardless of the choice of the metric $M$ and the complexity ($T^{(1)},T^{(2)}, \dots T^{(\omega)}$), by declaring a decoding failure if and only if the order of the noise realization is greater than $\omega$. These optimal performance serve as lower bounds on the FER results for practical SCFlip-$\omega$ decoders. We will further denote $\text{FER}_{\text{OA}\omega}$ the lower bound of the SCFlip-$\omega$ decoder.

\begin{figure}[!b]
\centering
\includegraphics[scale=0.42]{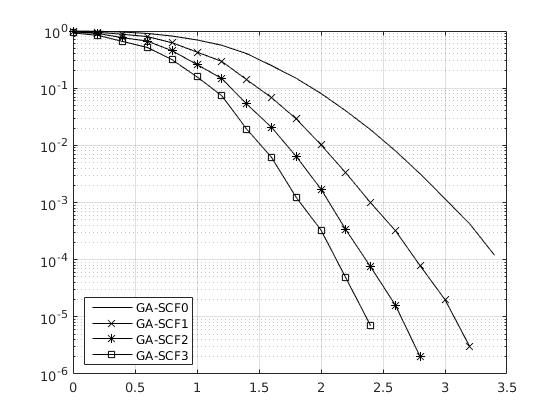} 
\caption{Lower bound of SCFlip-$\omega$ for a polar code $(N,K)=$(1024,512)}
\label{fig:perfsGA}
\end{figure}

Fig. \ref{fig:perfsGA} presents the lower-bounds of SCFlip-$\omega$ decoders with $\omega=\{1,2,3\}$ over BI-AWGN channel for a Polar code with parameters ($N,K+r$)=(1024,512+16). We also plot the performance of the SC decoder with $(N,K)=(1024,512)$. It can be seen that ideal SCFlip-$\omega$ decoders exhibit significant SNR gains compared to the SC decoder,  from $0.5$\,dB for the ideal SCFlip-1, to about $1$\,dB for the ideal SCFlip-2 decoder, at FER$=10^{-4}$. This ideal performance can be achieved with $T^{(\omega)}=\dbinom{K+r}{\omega}$ for any order $\omega$, assuming a perfect CRC, i.e. collision probability equal to 0. In practice, due to non perfect CRC, the probability of getting a CRC collision (hence an erroneous decoded message) increases with the number of decoding attempts. Therefore, optimizing the choice of flipping positions allows improving simultaneously the latency and the FER performance.

\subsection{Importance of SCFlip-1}

We define $P_{M}(\omega)$, the probability of not correcting a noise realization of order $\omega$ for a SCFlip-$\omega$ using a chosen metric $M$, a number $T^{(\omega)}$ of attempts, and assuming a perfect CRC. As a consequence, FER of SCFlip-$\omega$ can be lower bounded by:

\begin{equation}
\text{FER}_{\text{SCFlip}-\omega} \geq \text{FER}_{\text{OA}\omega}+\sum_{\omega'=1}^{\omega} P_M(\omega') \cdot D(\omega')
\end{equation}

\noindent where $D(\omega')$ denote the probability of the noise realization being of order $\omega'$ and $\text{FER}_{\text{OA}\omega}$ the lower bound determined by the OA-SC defined above. This is an inequality because of non perfect CRC. The sum in the right hand side of the above inequality is referred to as the {\em loss of order $\omega$}. It can be seen that this loss is incremental, so the loss of order $\omega$ will propagate at order $\omega+1$. In particular, to construct an SCFlip-2 decoder that closely approaches its theoretical lower bound $\text{FER}_{\text{OA}2}$, we would like loss of order 2 to be of same order of magnitude as $\text{FER}_{\text{OA}2}$. This implies that the loss of order 1 should also be of same order of magnitude as $\text{FER}_{\text{OA}2}$, and therefore order of magnitude smaller than $\text{FER}_{\text{OA}1}$. This condition is concretely materialised by a SCFlip-1 decoder matching its lower bound predicted by the OA-SC decoder.

\subsection{Complexity of SCFlip-$\omega$}

In addition to the FER performance, the performance of SCFlip-$\omega$ decoder is also characterized by its computational complexity. This computational complexity depends on the number of decoding attempts performed by the SCFlip-$\omega$ decoder in order to decode a given noise realization. Therefore, we denote by $N_c^{(ave)}$ and call the {\em normalized computational complexity} the average number of attempts. It is given by:
$$N_c^{(\text{ave})}=1+\text{FER}_{\text{SC}} \cdot T^{(\text{ave})} \underset{\text{SNR} \rightarrow +\infty}{\longrightarrow} 1,$$
\noindent where $T^{(\text{ave})}$ is the average number of decoding attempts and depends on the metric and the chosen values $T^{(\omega)}$. It is worth noticing that complexity of SCFlip tends to the one of the SC decoder when SNR tends to infinity.

\section{A New Metric for efficient scflip-$\omega$ Decoders}\label{sec:num4}

In \cite{afisiadis2014low}, the SCFlip decoder uses flipping positions which are ordered according to the absolute value of their LLRs. The criterion can be described as a metric $M(u_i), i \in \mathcal{I}$, defined by:
\begin{equation}
M(u_ i)=|\text{L}(u_i)|
\end{equation}
The flipping positions are those corresponding to the $T^{(1)}$ positions with the lowest $M(u_i)$. However, we point out that this metric is sub-optimal, because it does not take into account the sequential aspect of the SC decoder. Indeed, while a lower LLR absolute value indicates that the corresponding hard decision has a higher probability of being in error, it does not provide any information about the probability of being the first error that occurred during the sequential decoding process. To address this issue, we propose a new metric, which is aimed at identifying the first error that occurred during the sequential decoding process. The probability of $\hat{u}_k$ being the first error is given by:
$$\Pr(\hat{u}_k \ne u_k,\hat{u}_0^{k-1} =u_0^{k-1}  )=p_{\text{err}}(\hat{u}_k) \cdot \prod_{i=0}^{k-1} (1-p_{\text{err}}(\hat{u}_i))$$
where $p_{\text{err}}(\hat{u}_i)=\Pr (\hat{u}_i \ne u_i |Y, \hat{u}_0^{i-1}=u_0^{i-1})$. This probability cannot be computed in practice, as we have no guarantee that previous bits have been correctly decoded. Instead, we can compute the probability $\Pr (\hat{u}_i \ne u_i | Y,\hat{u}_0^{i-1})$ given  by (this follows from the definition of $L(u_i)$):
$$\Pr (\hat{u}_i \ne u_i |Y, \hat{u}_0^{i-1})=\frac{1}{1+\exp{(|\text{L}(u_i)|)}}$$
Note that if $u_i$ is a frozen bit, it cannot be in error, as the decoder always takes the right decision. Therefore, for frozen bits, the above probability is set to zero. 

To compute the probability of being the first error, we consider $\Pr (\hat{u}_i \ne u_i | \hat{u}_0^{i-1})$ as an approximation of $p_{\text{err}}(\hat{u}_i)$, and introduce a parameter $\alpha$ to compensate the approximation and which can be optimized by simulation. Thus, the proposed metric is given by:
\begin{definition}
Given a bit $u_k, k \in \mathcal{I}$, the metric associated to $u_k$ is defined by:
\begin{equation}
M_{\alpha}(u_k)=\frac{1}{1+\exp{(\alpha|\text{L}(u_k)|)}} \cdot \prod_{\underset{i \in \mathcal{I}}{i=0}}^{k-1} \left( \frac{1}{1+\exp{(-\alpha|\text{L}(u_i)|)}} \right)
\end{equation}

\noindent where $\alpha$ is a parameter to be optimized by simulation.
\end{definition}

We further define the equivalent logarithmic domain metric $M'_{\alpha}(u_k)=-\frac{1}{\alpha} \cdot \log(M_{\alpha}(u_k))$. It follows that:
\begin{align}
\begin{split}
M'_{\alpha}(u_k)&=|L(u_k)|+ \frac{1}{\alpha} \sum_{\underset{i \in \mathcal{I}}{i=0}}^{k-1} \left( 1+\exp(-\alpha \cdot |L(u_i)\ \right) \\
&= |L(u_k)|+ S_{\alpha}(u_0^k) \qquad \qquad \qquad \qquad \quad
\end{split}
\end{align}

The $S_{\alpha}(u_0^k)$ sum can be seen as a penalty added to $|L(u_k)|$, which take into consideration the sequential aspect of the SC decoding. Indeed, this term increases with increasing number and decreasing reliability of previously decoded bits, so that the last decoded bits are penalized compared to the metric (3). To understand the impact of the parameter $\alpha$, we consider the following limit cases. For $\alpha = 0$, the above metric becomes $M_0(u_k) = \frac{1}{2^{k_{\cal I}}}$, where $k_{\cal I}$ is the number of positions in ${\cal I}$ less than or equal to $k$.  Hence, the $M_0$ induced ordering corresponds to the usual decoding order. For $\alpha \longrightarrow +\infty$ it can be easily seen that $\lim\limits_{\alpha \rightarrow +\infty} S_{\alpha}(u_0^k)=0$, so that $M_{\infty}$ is equivalent to (3). In general, the use of the $M_\alpha$ metric  can be seen  an intermediate trade-off between the decoding order and the one given by the LLRs reliability (3). To find the best possible trade-off, we optimize the value of $\alpha$ by Monte-Carlo simulation. Note also that the optimized value should depend on the code used and the SNR. We provide an intuitive explanation of the behavior of the optimized alpha value as function of the SNR. Consider equation (5) for some fixed value of alpha. When the SNR goes to infinity, the term $S_{\alpha}(u_0^k)$ tends to 0 and becomes negligible compared to $|L(u_k)|$, and therefore the sequential characteristic of the decoder is no longer accounted for by the considered metric.  Consequently, it is expected that the optimal  value of alpha will increase with the SNR, so that to rebalance the contribution of the $S_{\alpha}(u_0^k)$ term to the value of the considered metric. To confirm this intuition, Table \ref{tab1} shows the optimized alpha values for a code $(N,K+r)= (1024,512+16)$ and several SNR values. As expected, it can be observed that the optimal alpha value increases with the SNR.

\begin{figure}[!t]
\centering
\includegraphics[scale=0.60]{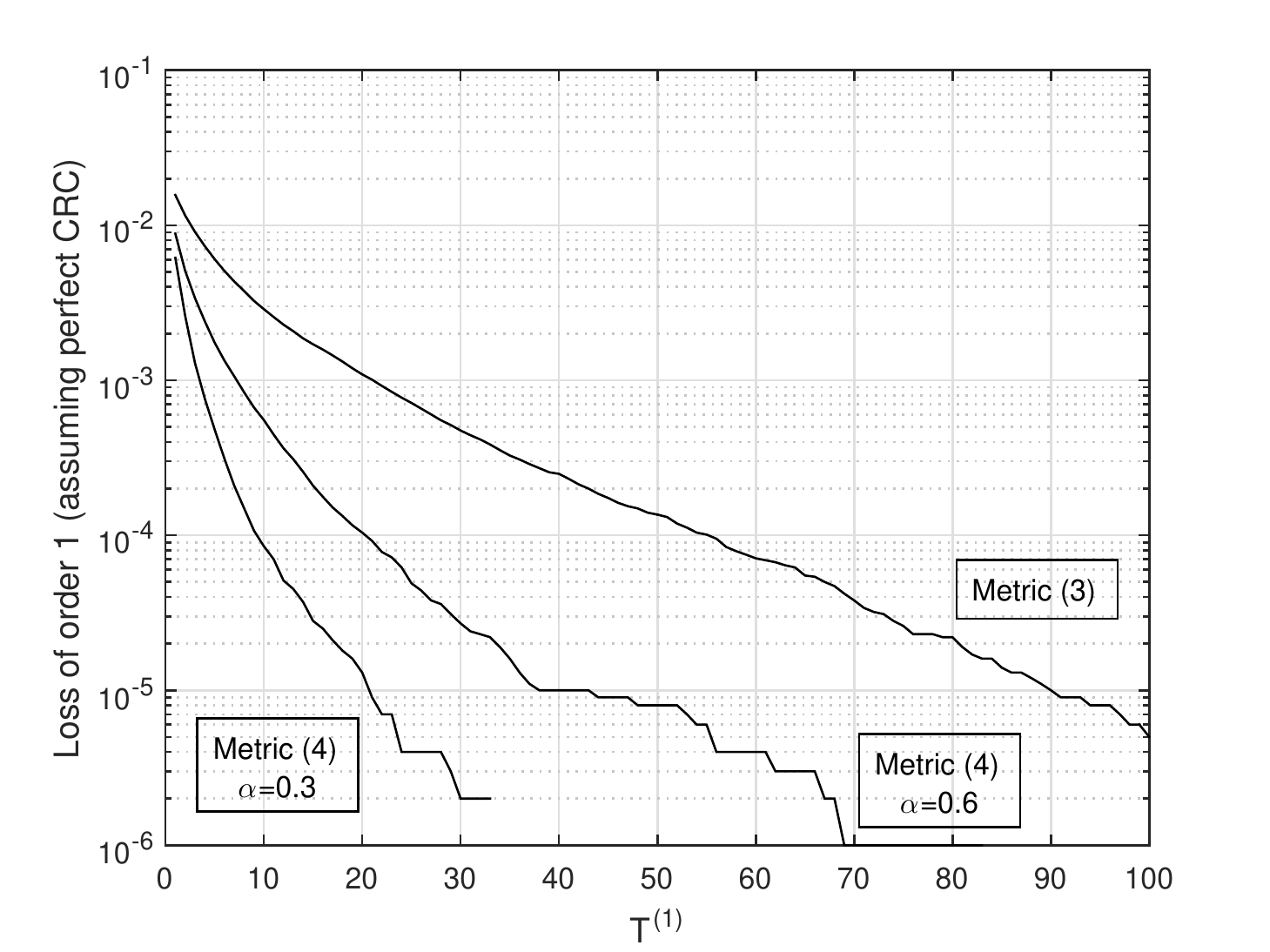} 
\caption{Loss of order 1 in function of $T^{(1)}$ for SNR=2.5dB}
\label{fig:ComparMetric}
\end{figure}

\begin{table}[!t]
\begin{center}
\caption{Optimized value of $\alpha$ for SCFlip-1 as function of the SNR for a code $(N,K+r)=(1024,512+16)$}
\label{tab1}
\begin{tabular}{|c|c|c|c|}
\hline
SNR (dB) & 1.5 & 2.5 & 3 \\
\hline
$\alpha_{opt}$ & 0.4 & 0.3 & 0.25 \\
\hline
\end{tabular}
\end{center}
\end{table}

We further consider a $(1024, 528)$ Polar code and consider only random noise realizations of order 1 at SNR\,$=2.5$\,dB. Fig.~2 plots the loss of order 1 ($P_M(1) \cdot D(1)$) assuming perfect CRC. It is calculated as the probability of the only \textit{channel-generated} error not being in the $\mathcal{L}_{\text{flip}}^{(1)}$ list generated by the \textit{FlipDetermine} procedure, as a function of the list size $T^{(1)}$, by using either the metric (3), or our proposed metric (4) with $\alpha = 0.3$ (optimized value) and $\alpha = 0.6$. According to Section III-C, we can determine the value of $T^{(1)}$ such that the loss of order 1 has the same order of magnitude as the theoretical lower bound of the SCFlip-2, $\text{FER}_{\text{OA2}}$. At SNR=2.5db, the SCFlip-2 lower bound is about $10^{-4}$, therefore, we need $D(1) \cdot P_M(1) < 10^{-4}$. This condition can easily be satisfied with the proposed metric (4), while we need $T^{(1)}$ much higher with the metric (3), which means in practice higher computational complexity. For next simulations, we choose $T^{(1)}=20$.

Let us now consider $\mathbf{Y}$ be a noise realization of order $\omega=2$ for a given code $\mathcal{C}(N,K,\mathcal{I})$ with the first CGE in position $u_{k1}$. We define the set $\mathcal{I}'= \{k \in \mathcal{I}, k>k1 \}$ of cardinality $K'<K$. Consider now the code $\mathcal{C}_2(N,K',\mathcal{I}')$. The order of the noise realization $\mathbf{Y}$ is only $\omega=1$ for the code $\mathcal{C}_2$ and therefore finding the second CGE is equivalent to decoding a noise realization of order 1 for the code $\mathcal{C}_2$. As a consequence, the metric $M_{\alpha}$ can be used also for SCFlip-2 by considering the set $\mathcal{I'}$ defined above instead of the set $\mathcal{I}$. However, the optimum value for $\alpha$ may be different. Numerical optimisation for $\alpha_2$ for a code $(1024,512+16)$ shows that the optimum value is $\alpha_2=0.5> \alpha_1=0.3$ at SNR=2.5dB. 

\section{Simulation Results}\label{sec:num5}

\begin{figure}[!tb]
\centering
\includegraphics[scale=0.60]{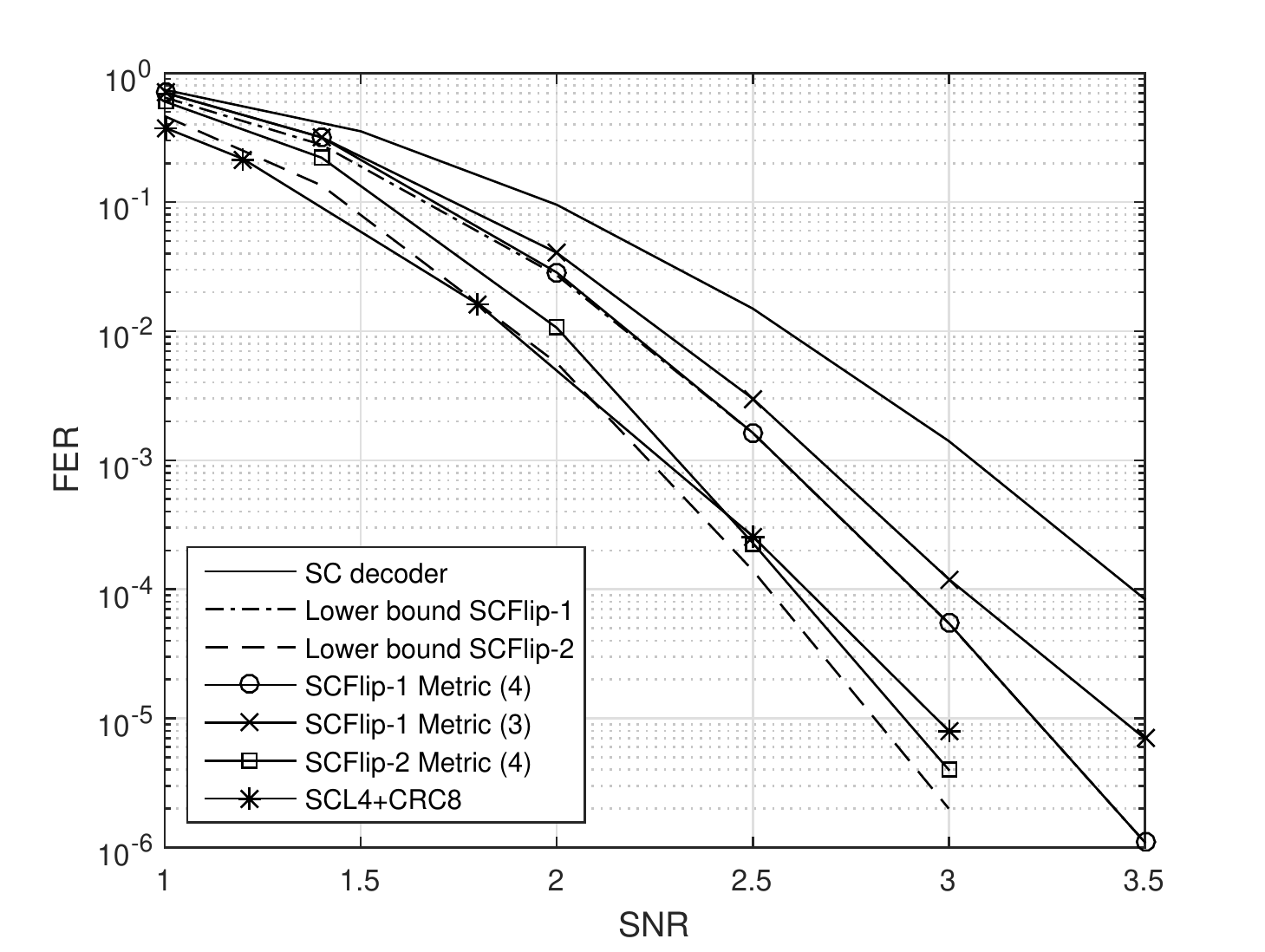} 
\caption{Performance of SCFlip-1 and SCFlip-2 decoders}
\label{fig:PerfSCF1}
\end{figure}

Throughout this section we consider transmission over a BI-AWGN channel, using a CRC-Polar code concatenation with parameters $N=1024$, $K=512$, and $r=16$. The $K+r$ positions for information and CRC bits, given by the set ${\cal I}$, are optimized for the SC decoder by Gaussian approximation as in \cite{trifonov2012efficient}. Also, this set is updated for each different value of the SNR. Concerning the CRC, we use a 16-bit CRC with generator polynomial   $g(x)=x^{16}+x^{15}+x^{2}+1$.

The FER performance of SCFlip-1 and SCFlip-2 decoders is shown in Fig. \ref{fig:PerfSCF1}. For SCFlip-1, the maximum number of flips is set to $T^{(1)} = 20$. We compare the SCFlip-1 decoder using the metric (3), with the one using our proposed metric (4) with $\alpha$=0.3, and the theoretical lower bound corresponding to an ideal SCFlip-1 decoder. While the performance gain compared to \cite{afisiadis2014low} is not impressive, it can be seen that our proposed metric closely approaches the theoretical lower bound. As the metric (4) exhibits a negligible loss of order 1, it can further be used to correct second order noise realizations (note that SCFlip-1 and SCFlip-2 using the metric (3) have nearly the same FER performance, since in this case the loss of order 1 is dominant). We further plot the FER performance of the proposed SCFlip-2 with $T^{(1)}=20, T^{(2,1)}=T^{(2,2)}=5, \alpha_1 = 0.3$, and $\alpha_2 = 0.5$, and compare with the CRC-aided SCL with $L$ = 4 and 16-bit CRC. The theoretical lower bound of SCFlip-2 decoder is also shown. It can be seen that the proposed SCFlip-2 closely approaches the theoretical lower bound, and exhibits nearly the same performance as the CRC-aided SCL decoder with $L=4$ at medium/high SNR.

\begin{figure}[!tb]
\includegraphics[scale=0.45]{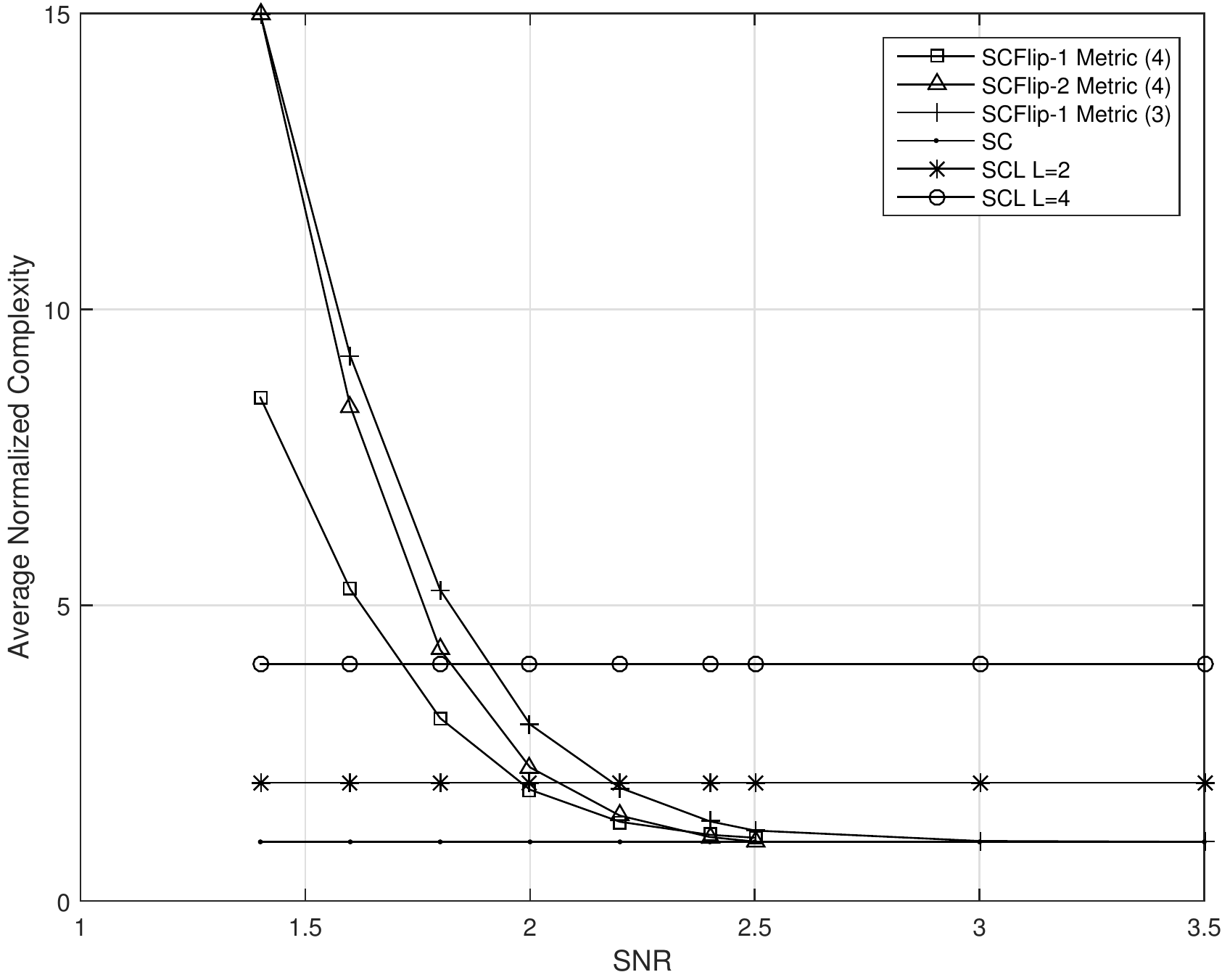} 
\caption{Average normalized computational complexity for SCFlip-1}
\label{fig:ComplSCF1}
\end{figure}

In Fig. \ref{fig:ComplSCF1}, we plot the average normalized complexity for SCFlip-1 with metric (3) and $T^{(1)}$=40, for SCFlip-1 with the proposed metric (4) with $T^{(1)}=20$, and SCFlip-2 with same parameters as in previous paragraph. Moreover, we also plot the normalized complexity of SC and SCL decoders with $L=\{2,4\}$. We note that the SCFlip-1 decoder exhibits similar FER performance when using our proposed metric (4) with $T^{(1)}=20$, or the metric (3) with $T^{(1)}=40$. We observe that the complexity of SCFlip is very high at low SNR, but converges quickly to the one of the SC decoder. For a SNR of 2.2dB, we have already a complexity lower than the one of the SCL with $L=2$. Moreover, one can see that the proposed metric allows reducing the average normalized complexity of the SCFlip-1 decoder by a factor of 2, as compared to the metric (3). Also, computational complexity of our proposed SCFlip-2 is even slightly better than SCFlip-1 with metric (3) and $T^{(1)}=40$, while it improves performance by 0.4dB at FER of $10^{-4}$.

\section{Conclusion}

In this paper, we first proposed an improvement of the SCFlip decoder of order-$1$, by introducing a new metric to determine the flipping positions, which takes into account the sequential aspect of the SC decoder. The proposed metric increases the ability of the SCFlip decoder to find the first error that occurred during the initial SC attempt, thus improving both decoding performance and computational complexity. Moreover, we have shown that the proposed metric allows closely approaching the theoretical lower bound corresponding to an ideal SCFlip decoder, and explained that this a necessary condition for building an effective SCFlip decoder of order 2.

We have further investigated an SCFlip decoder of order-2, which uses an analogous metric to determine the order-2 flip positions. We have shown that the SCFlip-2 decoder yields significant gains in terms of decoding performance, closely approaching the performance of the CRC-aided SC-List decoder with list size $L=4$, while having an average decoding complexity similar to that of the standard SC decoding at medium to high SNR.


\section*{Acknowledgment}

The research leading to these results received funding from the European Commission H2020 Programme, under grant agreement 671650 (mmMagic Project).

\bibliographystyle{IEEEtran}
\bibliography{biblio_database}



%

\end{document}